\documentclass[pdftex,twocolumn,epjc3]{svjour3}

\smartqed  % flush right qed marks, e.g. at end of proof

\RequirePackage{graphicx}
\RequirePackage[colorlinks,citecolor=blue,urlcolor=blue,linkcolor=blue]{hyperref}

\journalname{Eur. Phys. J. C}

\usepackage{lineno}
\usepackage{color}
\usepackage{ulem}
\usepackage{epstopdf}
\usepackage{subcaption}

\newcommand{\ecalRinn}{\ensuremath{R^{\mathrm{inner}}_{\mathrm{ECAL}}}}

\begin{document}

\title{Reconstruction and classification of tau lepton decays with ILD}

\author{T.H. Tran \thanksref{e1,addr1}
   \and
      V. Balagura \thanksref{addr1}
   \and
      V. Boudry   \thanksref{addr1}
   \and
      J.-C. Brient\thanksref{addr1}
   \and
      H. Videau   \thanksref{addr1}
}
\thankstext{e1}{e-mail: trong.hieu.tran@llr.in2p3.fr}
\institute{Laboratoire Leprince-Ringuet, \'Ecole polytechnique, CNRS/IN2P3,
   Universit\'e Paris-Saclay, 91128 Palaiseau, France \label{addr1}}

\maketitle
\begin{abstract}
  Tau-lepton decays with up to two $\pi^0$'s in the final state, $\tau^+ \to
  \pi^+ \bar{\nu}_\tau$, $\rho^+ (\pi^+\pi^0) \bar{\nu}_\tau$, $a^+_1
  (\pi^+\pi^0\pi^0) \bar{\nu}_\tau$, are used to study the performance of the
  barrel part of the silicon-tungsten electromagnetic calorimeter (Si-W ECAL)
  of International Large Detector (ILD) at the future $e^+-e^-$ International
  Linear Collider.  A correct reconstruction of the tau decay mode is crucial
  for constraining the tau spin state and measuring the Higgs boson CP state
  in $H\to \tau^+\tau^-$ decays. About 95\% of $\pi^+ \bar{\nu}_\tau$ and 90\%
  of $\rho^+\bar{\nu}_\tau$ and $a^+_1\bar{\nu}_\tau$ decays from $e^+e^-\to
  Z^0\to \tau^+\tau^-$ reaction at $e^\pm$-beam energy of 125~GeV are
  correctly reconstructed.  In a smaller ILD detector, with Si-W ECAL radius
  reduced by about 20\% these numbers degrade by at most 2\%. The $\pi^0$ mass
  resolution stays below 10\%. Since the failures in the tau-lepton
  reconstruction are mainly due to photons, the increase of the ILD magnetic
  field from 3.5~T to 4~T does not bring any significant improvement.
\end{abstract}
\vspace{3ex}

%\maketitle

\section{Introduction}\label{sec:intro}
The tau-lepton decays have been used at LEP~\cite{taupolLEP} and many other
experiments for precise tests of the Standard Model. The tau spin state can be
inferred from its decay mode and the momenta of all reconstructed decay
products~\cite{taupolarization}. This will be used at the future lepton
colliders, for example, to measure the CP (the product of charge conjugation
and parity symmetries) state of the Higgs boson decaying into a tau pair,
$H^0\to\tau^+\tau^-$.  In average, the tau decay products are more collimated
than the QCD jets of similar energy, and the separation of photon clusters in
the electromagnetic calorimeter (ECAL) is more difficult.  Incorrect
determination of the number of photons in the tau final state results in the
wrong reconstruction of the tau decay mode and spoils the spin state
measurement. This provides a thorough test of the ECAL performance.

In this paper we study the reconstruction and classification of $\tau^+$ decay
modes in the International Large Detector (ILD), a proposed detector concept
for the $e^+ - e^-$ International Linear Collider (ILC). We concentrate on the
reconstruction of three main tau decays with one charged pion\footnote{The
  inclusion of charge-conjugate states is implied in this article.} and up to
two $\pi^0$'s: $\tau^+ \to \pi^+ \bar{\nu}_\tau$, $\rho^+ \bar{\nu}_\tau$,
$a^+_1 \bar{\nu}_\tau$ followed by $\rho^+\to\pi^+\pi^0$,
$a_1^+\to\pi^+\pi^0\pi^0$ and $\pi^0\to\gamma\gamma$. For simplicity, we study
almost monochromatic tau-leptons from the decay of the virtual $Z^0$,
$e^+e^-\to Z^0\to \tau^+\tau^-$ at $e^\pm$-beam energy of 125~GeV. This
corresponds to 250~GeV centre-of-mass energy up to beamstrahlung and radiative
corrections. The tau-lepton momenta in this reaction are larger than, for
example, in $e^+e^-\to H^0Z^0$, $H^0\to\tau^+\tau^-$ at the same
centre-of-mass energy. Therefore, our results correspond to the Higgs boson
production at higher ILC energies.

Only the task of the correct classification of three $\tau^+$ decay modes
above is considered in this paper. Neither backgrounds are simulated (they are
expected to be small at ILC), nor any attempt is made to calculate $\tau^+$
spin state from the momenta of the reconstructed decay products.

The ILD detector is optimized for the particle flow algorithm approach
(PFA)~\cite{bib:pfa1,bib:pfa2}.  It combines the tracker systems with little
amount of material and highly granular, ``imaging'' calorimeters.  A complete
description of ILD is given in~\cite{bib:ildLOI,bib:ilcTDR}.  In this study
the ILD performance with a baseline design is simulated with
the software framework \textsc{Mokka}~\cite{bib:mokka} with a parametric
geometry description, based on the GEANT4 package~\cite{bib:geant4}.  The charged
pion from $\tau^+$ decay is reconstructed in ILD silicon vertex and tracker
detectors and the time projection chamber.  The photon clusters are
reconstructed in the ILD silicon-tungsten sampling electromagnetic calorimeter
(Si-W ECAL). We limit our analysis to the barrel part of ECAL, to avoid
complications in the region of barrel -- endcap overlap. The hadronic
calorimeter is not used.

Si-W ECAL has about 23 radiation lengths.  The silicon sensors have a
pixelization of $5\times5$~mm$^2$ and a thickness of 500~$\mu$m. Due to their
high cost and large area, Si-W ECAL is one of the most expensive subdetectors
in ILD.  Several studies of its cost optimization have been performed.
%In a Detailed Baseline Design report of ILD~\cite{bib:ildDBD}
In~\cite{bib:ilcTDR} it has been shown that a reduction of the number of ECAL
layers from 30 to 26 degrades the jet energy resolution by less than 5\% in a
jet energy range of 45--250~GeV.  In a more recent paper~\cite{bib:hieulcws},
it was demonstrated that a reduction of the ECAL radius by about 20\% degrades
the jet energy resolution by at most 8\% in the same energy range, while the
total price of ILD could be lowered by a factor of~1.5.

The current paper supplements the latter optimization study. In addition to
the baseline ILD design with the ECAL inner radius of 1843~mm, we present
results on the tau decay mode reconstruction in smaller ILD with the ECAL
radii of 1615 and 1450~mm. These numbers have been chosen according to the
size of the large industrially available silicon sensors in order to simplify
the final ECAL mechanical design.  The ECAL length is also reduced in order to
preserve the same radius-over-length aspect ratio. All other detector
parameters, like ECAL total and layer thicknesses, pixel size, gap between the
barrel and the endcaps, etc.\@, are unchanged.

The degradation of the jet and the track momentum resolution in smaller ILD
may be compensated by a higher magnetic field. Therefore, we also simulated
the ILD performance with the magnetic field increased from the nominal 3.5 to
4 T.

\section{Simulation and Reconstruction}\label{sec:simrec}

We used a sample of $e^+e^-\to Z^0 \to\tau^+\tau^-(\gamma)$ decays at 250~GeV
produced for the ILD ``Detailed Baseline Design'' report included in the ILC
``Technical Design Report''~\cite{bib:ilcTDR}. %~\cite{bib:ildDBD}.
The tau decay is generated by
the TAUOLA library~\cite{bib:tauola}.  Events with an initial state radiation
are removed at a generator level, but the final state radiation (FSR),
$Z^0\to\tau^+\tau^-\gamma$, is retained. FSR in the tau decays, $\tau^+\to
\pi^+\bar{\nu}_\tau\gamma,\ \rho^+\bar{\nu}_\tau\gamma,
\ a_1^+\bar{\nu}_\tau\gamma$, is also retained.  The tau branching fractions
of the three decays under study without FSR are listed in Table~\ref{tbl1}.
They cover 45\% of all tau decay modes and about 70\% of the hadronic decays.

\begin{table}
\centering
\begin{tabular}{lrccc|}\hline\hline
Decay mode                                          & Branching fraction [\%] \\\hline
$\pi^+\bar{\nu}_\tau$                               & $10.83\pm0.06$          \\
$\rho^+\bar{\nu}_\tau\;(\rho^+\to\pi^+\pi^0)$       & $25.52\pm0.09$          \\
$a_1^+\bar{\nu}_\tau\;(a_1^+\to\pi^+\pi^0\pi^0)$    & $9.30\pm0.11$           \\\hline\hline
\end{tabular}
\caption{Branching fractions of three tau decays under study with one charged
  and up to two neutral pions in the final
  state~\cite{bib:pdg}.}
\label{tbl1}
\end{table}

Each of the two $\tau$-leptons is reconstructed only inside the hemisphere
defined by an axis pointing in the tau direction in the $Z^0$ rest frame.  The
events with a photon conversion before ECAL are filtered out by requiring that
only one charged track is reconstructed in each hemisphere.

\subsection{The \textsl{\textsc{Garlic}} clustering algorithm}

The Gamma Reconstruction at a Linear Collider Experiment (\textsc{Garlic})
package~\cite{bib:MReinhard,Jeans:2012jj}, version 3.0.3, is used to find
photon clusters. The calibration constants for converting the energy deposited
in the silicon sensors to the full photon energy are determined using samples
of 10~GeV photons.  This is done separately for the baseline and two smaller
ILD ECAL models to take into account the possible variation of the overall
weight of non-instrumented ECAL zones.  The difference between the calibration
constants for all three models is less than 1\%.

\subsection{Photon identification}

The final identification of photon clusters is performed using not
\textsc{Garlic} package but a specialized algorithm described below. It is
developed for the purposes of this analysis.

FSR photons are suppressed by requiring the photon energy to be greater than
0.5~GeV.  This eliminates 70\% of FSR photons and only 6\% of photons from
$\tau^+\to\rho^+\bar{\nu}_\tau,\ a_1^+ \bar{\nu}_\tau$ decays.

To distinguish the genuine photons from fake clusters, for example,
originating from the charged $\pi^+$ ha\-dro\-nic showers, a boosted decision tree
(BDT) is used with the following input variables:
\begin{itemize}
\item the distance from the cluster barycentre to the $\pi^+$ track at the
  ECAL front surface,
\item the cluster depth computed from the cluster start found by the
  \textsc{Garlic} algorithm,
\item the cluster transverse fractal dimension defined as the logarithm of the
  number of hits in the cluster, $N_1$, divided by the number of hits when the
  pixels are grouped into larger pseudo-cells of $4\times4$ pixels, $N_4$:
\begin{equation}
FD_4 = \log(N_1/N_4)/\log(4).
\end{equation}
The fractal dimension, introduced in~\cite{bib:FD}, reflects the shower
density which is higher in electromagnetic showers.
\item The hit energy distribution mean and the standard deviation,
\item the fraction of energy deposited between 5 and 10 radiation lengths
  after the cluster start.
\end{itemize}

25\% of all tau decays are used to train BDT, which is then applied to the
remaining 75\% of statistics.  The photon clusters where more than 50\% of hit
energy is produced by the true Monte Carlo photon from the tau decay, are used
as a signal for a BDT training. The rest of the clusters in the same data
sample serves as background.

\begin{figure*}
\centering
\begin{subfigure}[t]{0.35\textwidth}
\includegraphics[width=\textwidth]{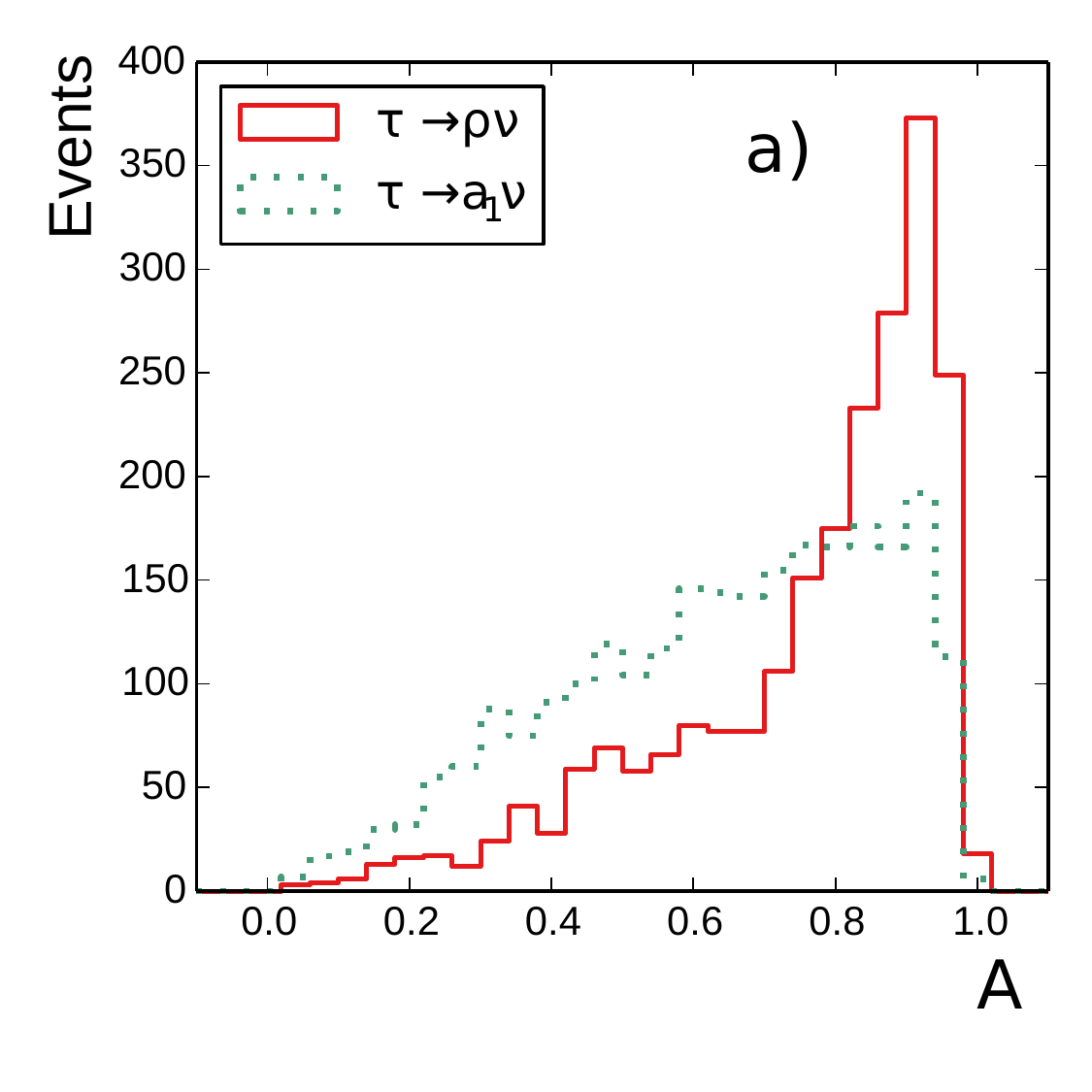}
\end{subfigure}
\hfil
\begin{subfigure}[t]{0.35\textwidth}
\includegraphics[width=\textwidth]{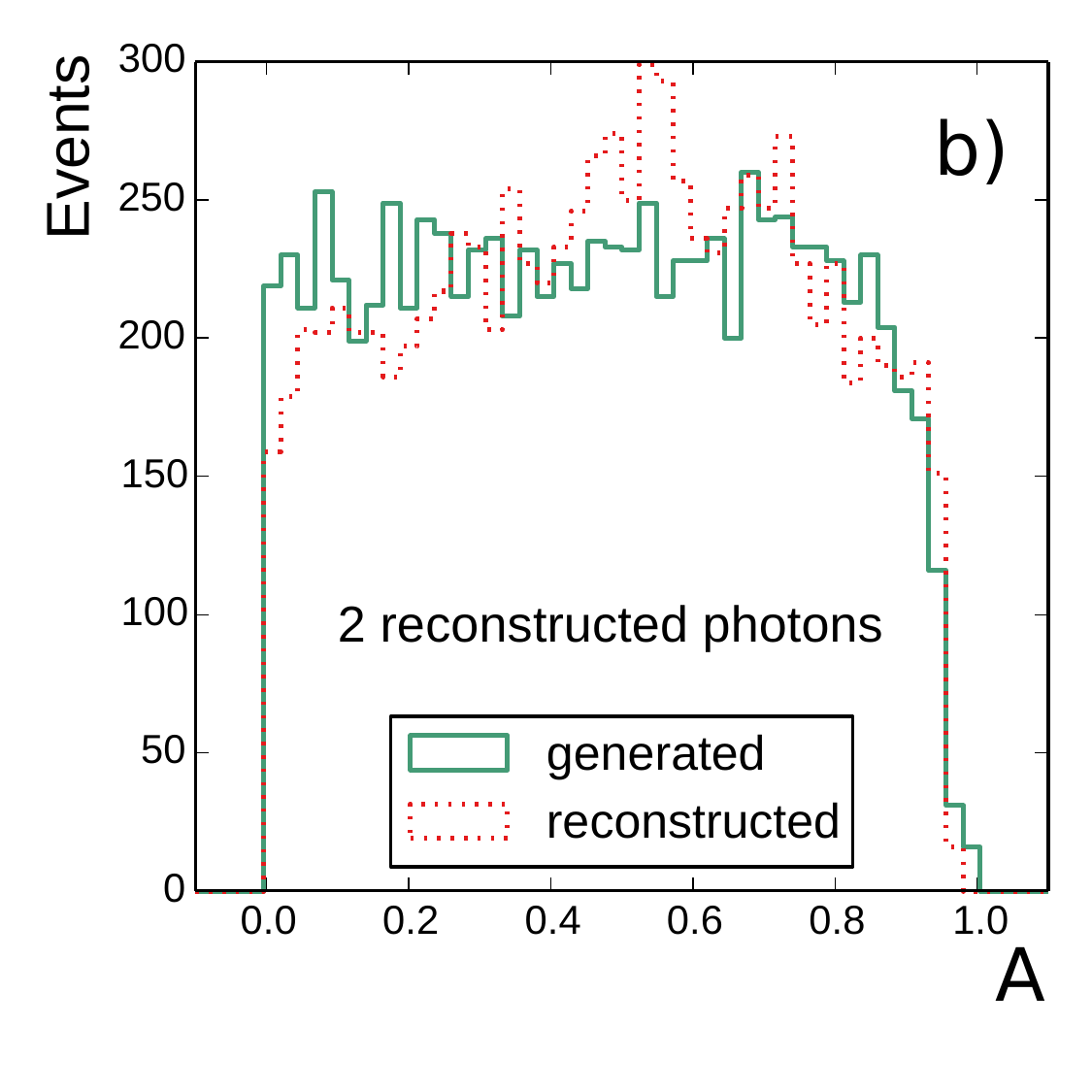}
\end{subfigure}
\caption{The photon energy asymmetry
  $A=(E^{\mathrm{max}}_\gamma-E^{\mathrm{min}}_\gamma)/(E^{\mathrm{max}}_\gamma+E^{\mathrm{min}}_\gamma)$
  in the case of three (a) and two (b) reconstructed photons.  Solid and
  dashed curves are for (a): two decay modes $\tau^+\to\rho^+\bar{\nu}_\tau$
  and $\tau^+\to a_1^+\bar{\nu}_\tau$, (b): for the generated (Monte Carlo
  true) and measured energies, respectively.}
\label{fig1}
\end{figure*}

The number of photons from $\tau^+$ decay without FSR should be
even. Sometimes, three photons are reconstructed, however.  This can happen if
one of the photon clusters from
$\tau^+\to\rho^+\bar{\nu}_\tau,\ \rho^+\to\pi^+\pi^0,\ \pi^0\to\gamma\gamma$
decays is split into two, or this decay is accompanied by the FSR photon. The
other possibility is the merging of two photon clusters out of four from $\tau^+\to
a_1^+\bar{\nu}_\tau,\ a_1^+\to\pi^+\pi^0\pi^0$ decay or loss of one photon.
In the first case of $\tau^+\to\rho^+\bar{\nu}_\tau$ decay, the photon with
the minimal energy is soft and the asymmetry between maximal and minimal
energies of three photons, defined as
\begin{equation}
A=\frac{E^{\mathrm{max}}_\gamma-E^{\mathrm{min}}_\gamma}{E^{\mathrm{max}}_\gamma+E^{\mathrm{min}}_\gamma},
\end{equation}
peaks at one. The distribution of $A$ for both $\tau^+\to\rho^+\bar{\nu}_\tau$
and $a_1^+\bar{\nu}_\tau$ is shown in Fig.~\ref{fig1}a. In this special case
of three photons, if $A>0.8$, the energy of the least energetic photon cluster
is added to the most energetic one and the cluster is removed from the
following analysis.

As a cross-check, the same asymmetry $A$ is plotted in Fig.~\ref{fig1}b in the
case of two reconstructed photons. It is dominated by
$\tau^+\to\rho^+\bar{\nu}_\tau$ decays.  Since $\pi^0$ has no spin, its decay
products should have a flat asymmetry distribution, ranging from zero to one
as the photon is massless. This is really the case for the true Monte Carlo
photon energies up to acceptance and detector cut-off effects at one (the
solid curve in Fig.~\ref{fig1}b). The dashed curve, obtained after
reconstruction, shows, however, that the clustering algorithm tends to favor
an asymmetric energy partition between the two close photons.

\subsection{Classification of tau decay modes}
\label{sec23}

To distinguish three $\tau^+$ decay modes, the second BDT, with gradient
boost is used with the following input variables:
\begin{itemize}
\item the invariant mass of all reconstructed $\tau^+$ decay products,
  $M_{\mathrm{reco}}$,
\item the number of reconstructed photons and
\item their total invariant mass,
\item the energy of the photons and
\item their distance from the reconstructed track at the ECAL front surface.
\end{itemize}
For every decay mode a BDT is trained on 25\% of statistics using this decay
as a signal and the other two as a background. Three classifiers, obtained in
this way (one for every mode), are applied to the remaining 75\% of events.

If $M_{\mathrm{reco}}$ is larger than the tau-lepton mass, the event is
rejected, except if it is classified as $\pi^+\bar{\nu}_\tau$ with one
reconstructed photon. In the latter case it is assumed that the mass beyond
the kinematical limit is due to an external FSR photon and the event is
accepted.

A small fraction of events is accepted at the same time by two classifiers,
mainly when the final state with three reconstructed photons is compatible
with both $\rho^+\bar{\nu}_\tau$ and $a_1^+\bar{\nu}_\tau$ decays.  Then, the
final decision is made based on $M_{\mathrm{reco}}$: $\rho^+\bar{\nu}_\tau$ is
preferred over $a_1^+\bar{\nu}_\tau$ if $M_{\mathrm{reco}}< 0.85$~GeV and vice
versa. For events accepted at the same time by two other classifiers
($\pi^+\bar{\nu}_\tau$ and $\rho^+\bar{\nu}_\tau$ or $\pi^+\bar{\nu}_\tau$ and
$a_1^+\bar{\nu}_\tau$), the preference is made in favor of the heavier
particle in the final state.  No event passes all three classifiers.

\section{Results and conclusions}

\begin{figure*}
\centering
\includegraphics[width=0.75\textwidth]{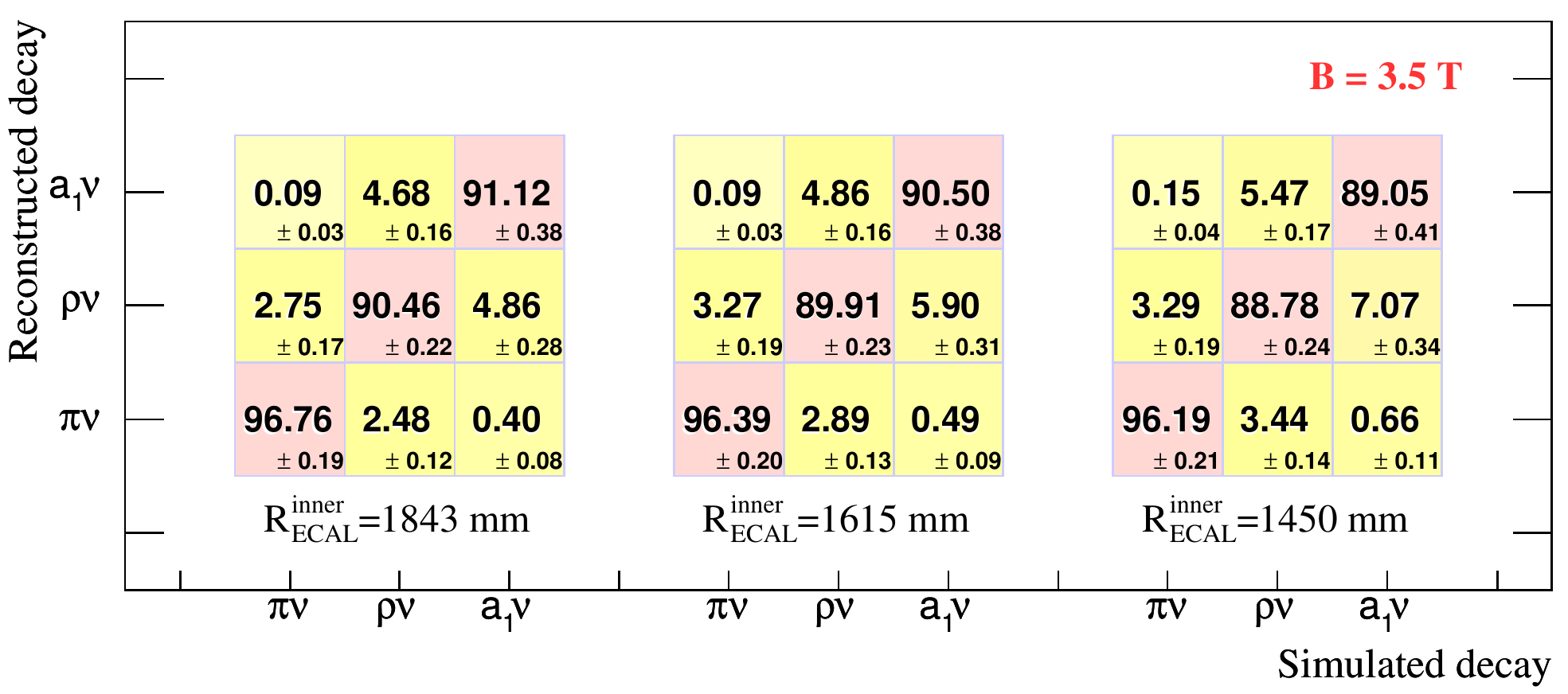}
\caption{The probability of the correct and wrong reconstruction of the tau
  decay modes $\tau^+\to\pi^+\bar{\nu}_\tau$, $\rho^+\bar{\nu}_\tau$ and
  $a_1^+\bar{\nu}_\tau$ in percent, obtained for Si-W ECAL inner radius (from
  left to right) $\ecalRinn=1843$ (baseline), 1615 and 1450~mm and with the
  nominal magnetic field of $3.5$~T. Only the statistical uncertainties are
  shown.}
\label{fig2}
\end{figure*}

Fig~\ref{fig2} summarizes the probability of the correct reconstruction of tau
decay modes $\tau^+\to\pi^+\bar{\nu}_\tau$, $\rho^+\bar{\nu}_\tau$ and
$a_1^+\bar{\nu}_\tau$ in the absence of backgrounds in the barrel Si-W ECAL
region. The results are presented for the ILD baseline design with the inner
ECAL radius $\ecalRinn=1843$~mm and for two models of smaller ILD with
$\ecalRinn$ $=1615$ and 1450~mm. The probability does not include the efficiency
of charged $\pi^+$ reconstruction, which is taken to be 100\%. The
off-diagonal values correspond to the probabilities of wrong
reconstruction. For some events the decay mode could not be determined (either
$M_{\mathrm{reco}}$ is larger than the tau-lepton mass or the event is not
accepted by any classifier, as explained in Sec.~\ref{sec23}), therefore the
sum of the probabilities in columns and rows do not reach 100\%.  The
reconstruction of two tau-leptons in $Z^0\to\tau^+\tau^-(\gamma)$ events is
performed independently, so that a failure to reconstruct one does not affect
the other. The errors are purely statistical.

The correct reconstruction probability is close to 90\% for
$\rho^+\bar{\nu}_\tau$ and $a_1^+\bar{\nu}_\tau$ and above 95\% for
$\pi^+\bar{\nu}_\tau$ decays.  For the smallest ILD with $\ecalRinn=1450$~mm
the degradation of the probabilities compared to the baseline design is less
than 2\%.

\begin{figure*}
\centering
\includegraphics[width=.58\textwidth]{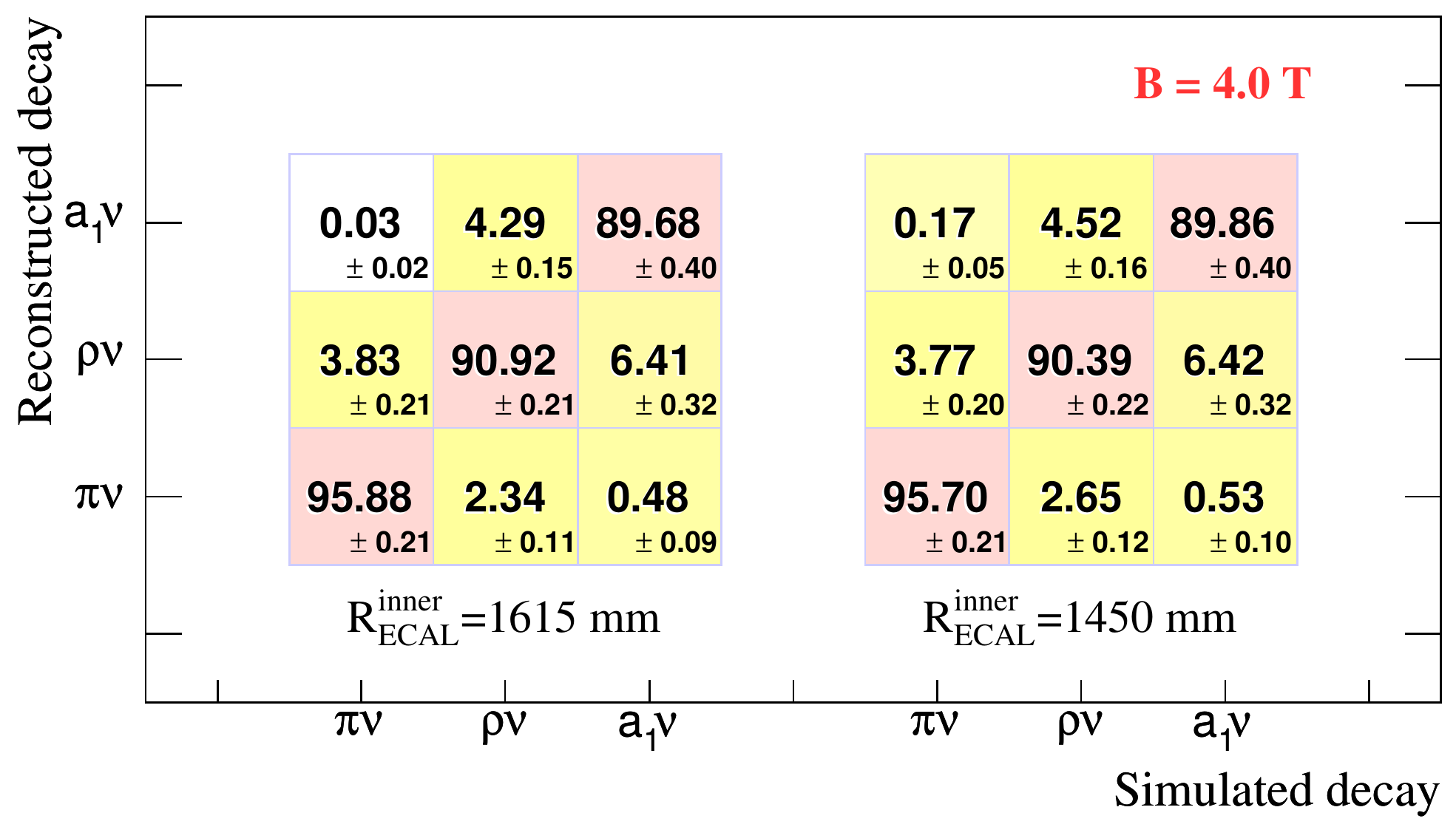}
\caption{The same as in Fig.~\ref{fig2}, but for the magnetic field of 4~T.}
\label{fig3}
\end{figure*}

For smaller ILD model there is an option to compensate the slightly degraded
jet and track momentum resolution by a higher magnetic field. This increases
the bending of the charged tracks and hadron-hadron and hadron-photon
separation in the jets.  Therefore, we also simulated the ILD performance with
the magnetic field increased from the nominal 3.5 to 4~T.  The corresponding
reconstruction probabilities are shown in Fig~\ref{fig3}.  The improvement is
marginal and is less than 1\%. This demonstrates that the failure rate is
dominated by the photon reconstruction which is almost independent of the
magnetic field.

\begin{figure*}
\centering
\begin{subfigure}[t]{0.35\textwidth}
\includegraphics[width=\textwidth]{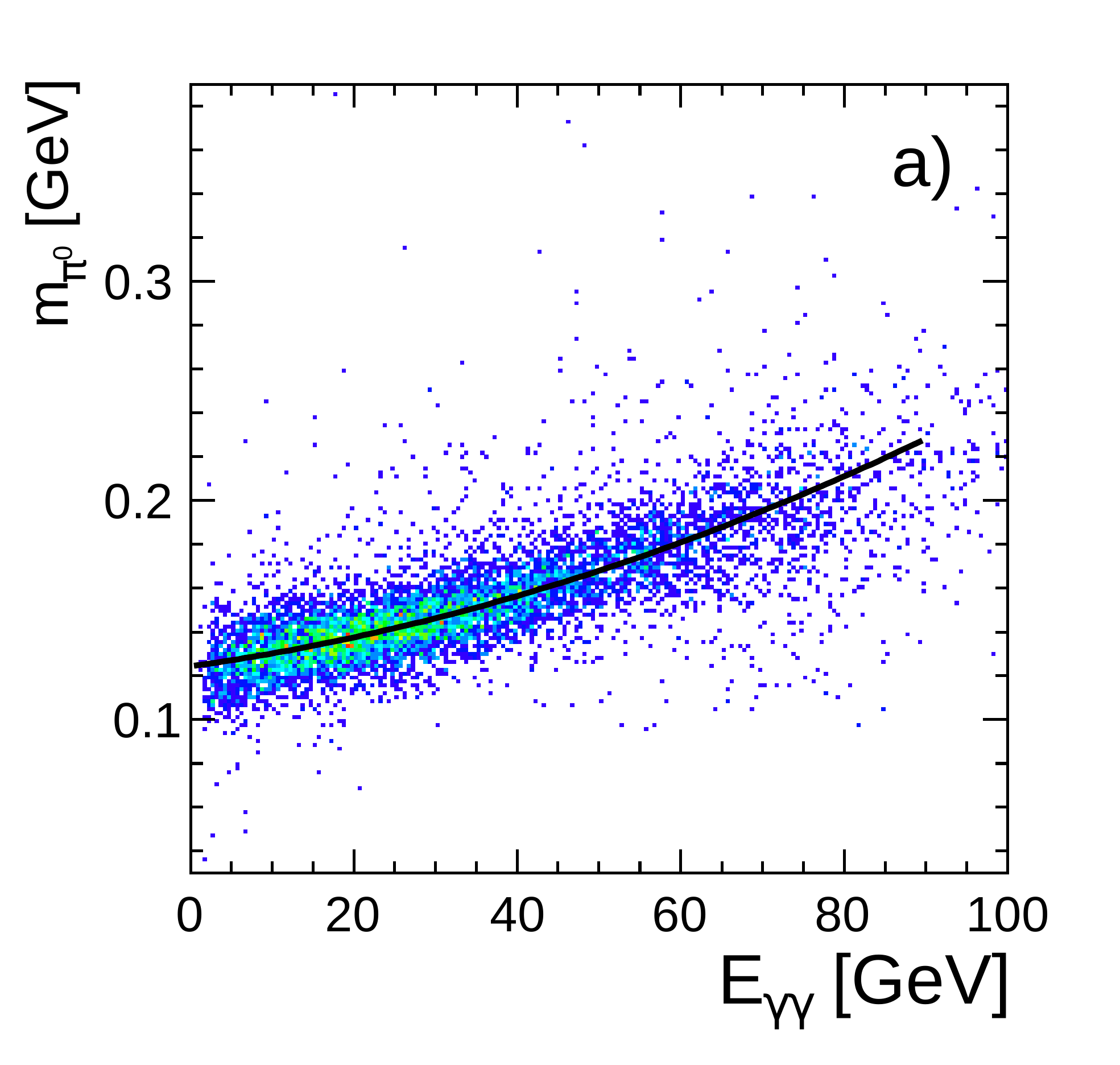}
\end{subfigure}
\hfil
\begin{subfigure}[t]{0.35\textwidth}
\includegraphics[width=\textwidth]{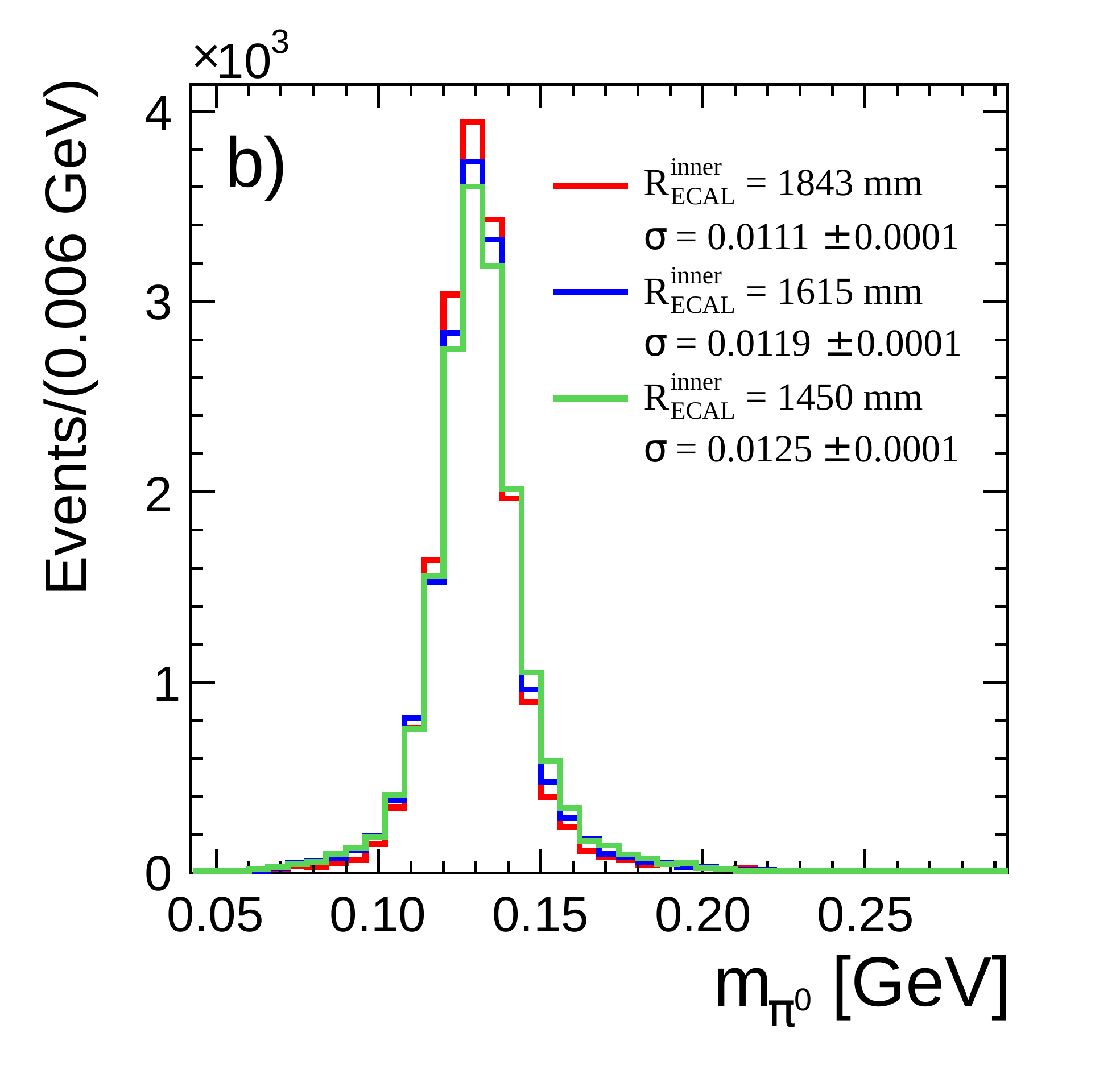}
\end{subfigure}
\caption{Left: the dependence of the reconstructed mass of $\pi^0$ on its
  energy before the correction. The black curve shows the parabolic
  fit. Right: the corrected $\pi^0$ mass and the resolution for ILD with three
  different ECAL radii.}
\label{fig4}
\end{figure*}

The photon clusters from the decays of energetic $\pi^0$'s often overlap and
this degrades the $\pi^0$ mass resolution. The effect is stronger for smaller
ILD ECAL with less photon separation. We study this effect using a sample with
two reconstructed photons dominated by $\rho^+\bar{\nu}_\tau$
($\rho^+\to\pi^+\pi^0$) decays and therefore having more energetic
$\pi^0$. Fig.~\ref{fig4}a shows that the reconstructed invariant mass of
$\pi^0$ increases with its energy. This is explained by the fact that the
\textsc{Garlic} algorithm splits the overlapping showers geometrically into
two disjoint groups of neighboring hits, neglecting the fact that some part of
the cluster energy can propagate into the area of another cluster.  This
overestimates the distance between the cluster barycenters and the opening angle
between the photons. The distribution in Fig.~\ref{fig4}a is fit to a
parabolic function and the corresponding correction is applied to recover the
nominal $\pi^0$ mass at all energies. The resulting mass resolution, shown in
Fig.~\ref{fig4}b, has a symmetric shape. The relative $\pi^0$ mass resolution
is below 10\% for all considered ILD models. The difference between the
nominal and the smallest ILD models is about 1\%.

A proper reconstruction of the tau decay modes together with a good momentum
measurement of the decay products is mandatory to reconstruct the tau spin
states in the measurement of the Higgs boson CP state in the decay
$H^0\to\tau^+\tau^-$.  The high probability of the correct decay mode
reconstruction demonstrated in this paper shows the full potential of ILD even
with the reduced size for such a measurement.

\begin{acknowledgements}
  The authors would like to thank Dr. Daniel Jeans, University of Tokyo for
  the help with the \textsc{Garlic} package used in this study. This work is
  funded by the Physics Department of \'Ecole polytechnique, Palaiseau, France
  and the ``Physique des deux Infinis et des Origines'' (P2IO) program.
\end{acknowledgements}

\bibliographystyle{spphys}
%\bibliography{ildopt}
%\begin{thebibliography}{9}   % Use for  1-9  references

\end{document}